\documentclass[longauth]{aa}
\usepackage{graphicx}
\usepackage{txfonts}
\usepackage{xcolor}
\usepackage{hyperref}
\hypersetup{colorlinks, linkcolor=blue, citecolor=blue, urlcolor=blue}
\usepackage{amsmath}
\usepackage{ulem}
\usepackage{soul}
\usepackage{enumitem}

\newcommand{\muvec}{\mbox{\boldmath $\mu$}}

\newcommand{\te}{t_{\rm E}}
\newcommand{\thetae}{\theta_{\rm E}}

\newcommand{\pie}{\pi_{\rm E}}

\newcommand{\dl}{D_{\rm L}}

\definecolor{brown}{rgb}{0.59, 0.29, 0.0}
\definecolor{darkgreen}{rgb}{0.0, 0.42, 0.24}
\definecolor{darkblue}{rgb}{0.01, 0.31, 0.59}
\definecolor{darkblue}{rgb}{0.0, 0.25, 0.42}
\definecolor{blue}{rgb}{0.0,0.0,1.0}
\definecolor{green}{rgb}{0.0,1.0,0.0}

\begin{document}

\title{Four sub-Jovian-mass planets detected by high-cadence microlensing surveys}

\author{
     Cheongho~Han\inst{01} 
\and Doeon~Kim\inst{01}
\and Andrew~Gould\inst{02,03}      
\and Andrzej~Udalski\inst{04} 
\and Ian A. Bond\inst{05}
\and Valerio Bozza\inst{06}
\and Youn~Kil~Jung\inst{07} 
\\
(Leading authors)\\
     Michael~D.~Albrow\inst{08}   
\and Sun-Ju~Chung\inst{07}     
\and Kyu-Ha~Hwang\inst{07} 
\and Yoon-Hyun~Ryu\inst{07} 
\and In-Gu~Shin\inst{07} 
\and Yossi~Shvartzvald\inst{09}   
\and Jennifer~C.~Yee\inst{10}   
\and Weicheng~Zang\inst{11}     
\and Sang-Mok~Cha\inst{07,12} 
\and Dong-Jin~Kim\inst{07} 
\and Seung-Lee~Kim\inst{07} 
\and Chung-Uk~Lee\inst{07} 
\and Dong-Joo~Lee\inst{07} 
\and Yongseok~Lee\inst{07,13} 
\and Byeong-Gon~Park\inst{07,13} 
\and Richard~W.~Pogge\inst{03}
\\
(The KMTNet Collaboration),\\
     Przemek~Mr{\'o}z\inst{04} 
\and Micha{\l}~K.~Szyma{\'n}ski\inst{04}
\and Jan~Skowron\inst{04}
\and Rados{\l}aw~Poleski\inst{04} 
\and Igor~Soszy{\'n}ski\inst{04}
\and Pawe{\l}~Pietrukowicz\inst{04}
\and Szymon~Koz{\l}owski\inst{04} 
\and Krzysztof~Ulaczyk\inst{14}
\and Krzysztof~A.~Rybicki\inst{04}
\and Patryk~Iwanek\inst{04}
\\
(The OGLE Collaboration),\\
     Fumio~Abe\inst{15}
\and Richard~K.~Barry\inst{15} 
\and David~P.~Bennett\inst{17,18} 
\and Aparna~Bhattacharya\inst{17,18}
\and Hirosane~Fujii\inst{15}
\and Akihiko~Fukui\inst{19,20} 
\and Yuki~Hirao\inst{16,17,21}
\and Yoshitaka~Itow\inst{15}
\and Rintaro~Kirikawa\inst{21} 
\and Naoki~Koshimoto\inst{21} 
\and Iona~Kondo\inst{21} 
\and Yutaka~Matsubara\inst{15}
\and Sho~Matsumoto\inst{21}
\and Shota~Miyazaki\inst{21} 
\and Yasushi~Muraki\inst{21} 
\and Greg~Olmschenk\inst{23}
\and Arisa~Okamura\inst{21} 
\and Cl\'ement~Ranc\inst{24} 
\and Nicholas~J.~Rattenbury\inst{25} 
\and Yuki~Satoh\inst{21}
\and Stela~Ishitani Silva\inst{17,26}
\and Takahiro~Sumi\inst{21}
\and Daisuke~Suzuki\inst{27}
\and Taiga~Toda\inst{21}
\and Paul~J.~Tristram\inst{28} 
\and Aikaterini~Vandorou\inst{23} 
\and Hibiki~Yama\inst{21}
\\
(The MOA Collaboration)\\
}

\institute{
     Department of Physics, Chungbuk National University, Cheongju 28644, Republic of Korea  \\ \email{cheongho@astroph.chungbuk.ac.kr}                  
\and Max Planck Institute for Astronomy, K\"onigstuhl 17, D-69117 Heidelberg, Germany                                                                    
\and Department of Astronomy, The Ohio State University, 140 W. 18th Ave., Columbus, OH 43210, USA                                                       
\and Astronomical Observatory, University of Warsaw, Al.~Ujazdowskie 4, 00-478 Warszawa, Poland                                                          
\and Institute of Information and Mathematical Sciences, Massey University, Private Bag 102-904, North Shore Mail Centre, Auckland, New Zealand          
\and Dipartimento di Fisica `E.R. Caianiello', Universit a di Salerno, Via Giovanni Paolo 132, Fisciano I-84084, Italy                                   
\and Korea Astronomy and Space Science Institute, Daejon 34055, Republic of Korea                                                                        
\and University of Canterbury, Department of Physics and Astronomy, Private Bag 4800, Christchurch 8020, New Zealand                                     
\and Department of Particle Physics and Astrophysics, Weizmann Institute of Science, Rehovot 76100, Israel                                               
\and Center for Astrophysics $|$ Harvard \& Smithsonian 60 Garden St., Cambridge, MA 02138, USA                                                          
\and Department of Astronomy and Tsinghua Centre for Astrophysics, Tsinghua University, Beijing 100084, China                                            
\and School of Space Research, Kyung Hee University, Yongin, Kyeonggi 17104, Republic of Korea                                                           
\and Korea University of Science and Technology, 217 Gajeong-ro, Yuseong-gu, Daejeon, 34113, Republic of Korea                                           
\and Department of Physics, University of Warwick, Gibbet Hill Road, Coventry, CV4 7AL, UK                                                               
\and Institute for Space-Earth Environmental Research, Nagoya University, Nagoya 464-8601, Japan                                                         
\and Astrophysics Science Division, NASA/Goddard Space Flight Center, Greenbelt, MD20771, USA                                                            
\and Laboratory for Exoplanets and Stellar Astrophysics, NASA / Goddard Space Flight Center, Greenbelt, MD 20771, USA                                    
\and Department of Astronomy, University of Maryland, College Park, MD 20742, USA                                                                        
\and Department of Earth and Planetary Science, Graduate School of Science, The University of Tokyo, 7-3-1 Hongo, Bunkyo-ku, Tokyo                       
\and Instituto de Astrof\'isica de Canarias, V\'ia L\'actea s/n, E-38205 La Laguna, Tenerife, Spain                                                      
\and Department of Earth and Space Science, Graduate School of Science, Osaka University, 1-1 Machikaneyama, Toyonaka, Osaka 560-0043, Japan             
\and Department of Astronomy, Graduate School of Science, The University of Tokyo, 7-3-1 Hongo, Bunkyo-ku, Tokyo 113-0033, Japan                         
\and Code 667, NASA Goddard Space Flight Center, Greenbelt, MD 20771, USA                                                                                
\and Zentrum f{\"u}r Astronomie der Universit{\"a}t Heidelberg, Astronomisches Rechen-Institut, M{\"o}nchhofstr.\ 12-14, 69120 Heidelberg, Germany       
\and Department of Physics, University of Auckland, Private Bag 92019, Auckland, New Zealand                                                             
\and Department of Physics, The Catholic University of America, Washington, DC 20064, USA                                                                
\and Institute of Space and Astronautical Science, Japan Aerospace Exploration Agency, 3-1-1 Yoshinodai, Chuo, Sagamihara, Kanagawa, 252-5210, Japan     
\and University of Canterbury Mt. John Observatory, P.O. Box 56, Lake Tekapo 8770, New Zealand                                                           
}
\date{Received ; accepted}

\abstract
{}
{
With the aim of finding short-term planetary signals,  we investigated the data 
collected from the high-cadence microlensing surveys.
}
{
From this investigation, we found four planetary systems with low planet-to-host mass ratios,
including OGLE-2017-BLG-1691L, KMT-2021-BLG-0320L, KMT-2021-BLG-1303L, and KMT-2021-BLG-1554L.
Despite the short durations, ranging from a few hours to a couple of days, the planetary signals 
were clearly detected by the combined data of the lensing surveys. It is found that three 
of the planetary systems have mass ratios of the order of $10^{-4}$ and the other has a mass ratio 
slightly greater than $10^{-3}$.
}
{
The estimated masses indicate that all discovered planets have sub-Jovian masses. The planet 
masses of KMT-2021-BLG-0320Lb, KMT-2021-BLG-1303Lb, and KMT-2021-BLG-1554Lb correspond to $\sim 
0.10$, $\sim 0.38$, and $\sim 0.12$ times of the mass of the Jupiter, and the mass of OGLE-2017-BLG-1691Lb 
corresponds to that of the Uranus.  The estimated mass of the planet host KMT-2021-BLG-1554L,  
$M_{\rm host}\sim 0.08~M_\odot$, corresponds to the boundary between a star and a brown dwarf.  
Besides this system, the host stars of the other planetary systems are low-mass stars with masses 
in the range of $\sim [0.3$--$0.6]~M_\odot$.  The discoveries of the planets well demonstrate the 
capability of the current high-cadence microlensing surveys in detecting low-mass planets. 
}
{}


\maketitle

\section{Introduction}\label{sec:one}

During the 2010s, microlensing entered an era of high-cadence surveys with the instrumental
upgrade of previously established experiments and the participation of a new experiment. 
The new era started when the Microlensing Observations in Astrophysics survey 
\citep[MOA:][]{Bond2001}, which had previously carried out a lensing survey using a 0.61~m telescope 
and a camera with a 1.3~deg$^{2}$ field of view (FOV) during the early phase, launched its second 
phase experiment with the employment of a 1.8~m telescope equipped with a wide-field camera yielding
a 2.2~deg$^{2}$ FOV. Since its first operation in 1992, the Optical Gravitational Lensing Experiment
(OGLE) has been upgraded multiple times, and it is in its fourth phase \citep[OGLE-IV:][]{Udalski2015} 
with the employment of a 1.3~m telescope and a camera with a 1.4~deg$^{2}$ FOV. The Korea
Microlensing Telescope Network \citep[KMTNet:][]{Kim2016}, which was launched in 2016, is being
carried out with the use of three globally distributed 1.6~m telescopes, each of which is
mounted with a wide-field detector providing a 4~deg$^{2}$ FOV. With the use of wide-field cameras
mounted on multiple telescopes, the current lensing surveys can monitor lensing events with a
dramatically enhanced observational cadence.

\begin{table*}[t]
\small
\caption{Coordinates, fields, alert dates, and baseline magnitudes\label{table:one}}
\begin{tabular}{llllll}
\hline\hline
\multicolumn{1}{c}{Event}                     &
\multicolumn{1}{c}{(RA, decl)$_{\rm J2000}$}  &
\multicolumn{1}{c}{$(l, b)$}                  &
\multicolumn{1}{c}{Field}                     &
\multicolumn{1}{c}{Alert date}                &
\multicolumn{1}{c}{$I_{\rm base}$ (mag)}      \\
\hline
 OGLE-2017-BLG-1691  & (17:34:22.52, -29:17:05.39)    &  $(-1^\circ\hskip-2pt .601,  1^\circ\hskip-2pt .895)$   &  OGLE (BLG654.31)     &  2017 Sep 6  &  19.90  \\ 
 (KMT-2017-BLG-0752) &                                &                                                         &  KMT (BLG14)          &  postseason  &         \\ 
\hline                                                                                                                                       
 KMT-2021-BLG-0320   & (17:57:33.15, -30:30:14.18)    &  $( 2^\circ\hskip-2pt .095, -4^\circ\hskip-2pt .299)$   &  KMT (BLG01, BLG41)   &  2021 Apr 9  &  20.08  \\ 
\hline                                                                                                                                    
 KMT-2021-BLG-1303   & (18:07:27.33, -29:16:53.69)    &  $(-0^\circ\hskip-2pt .027, -3^\circ\hskip-2pt .028)$   &  KMT (BLG04)          &  2021 Jun 14 &  19.67  \\ 
 (MOA-2021-BLG-182)  &                                &                                                         &  MOA (gb13)           &  2021 Jun 17 &         \\ 
\hline                                                                                                                                    
 KMT-2021-BLG-1554   & (17:51:12.82, -31:51:51.70)    &  $(-1^\circ\hskip-2pt .888, -2^\circ\hskip-2pt .543)$   &  KMT (BLG01, BLG41)   &  2021 Jul 1  &  22.39  \\ 
\hline                                             
\end{tabular}
\tablefoot{ ${\rm HJD}^\prime = {\rm HJD}- 2450000$.  }
\end{table*}

\begin{figure}[t]
\includegraphics[width=\columnwidth]{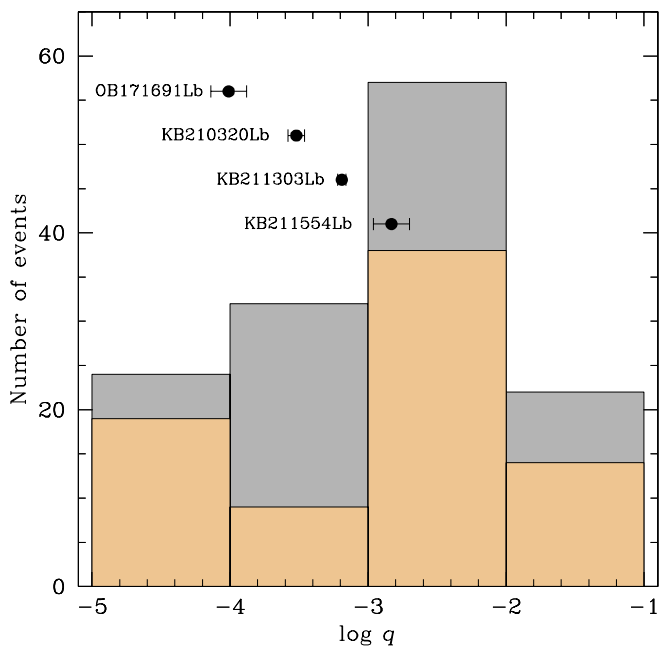}
\caption{
Histogram of published microlensing planets as a function of the planet-to-host mass ratio. The 
histogram of planets detected with the data from KMTNet survey is separately marked in brown. 
The solid dots with error bars indicate the four planets reported in this work.
}
\label{fig:one}
\end{figure}

The greatly increased observational cadence of the lensing surveys has brought out important 
changes not only in the observational strategy but also in the outcome of planetary microlensing 
searches.  First, being able to resolve short-lasting planetary signals from survey observations, 
the high-cadence surveys can function without the survey+followup experiment mode, in which 
low-cadence surveys mainly detect and alert lensing events and followup groups densely observe 
the alerted events.  Second, the number of microlensing planets has substantially increased 
with the operation of the high-cadence surveys.  Among the total 135 microlensing planets with 
measured mass ratios\footnote{The sample includes 119 planets listed in the NASA Exoplanet 
Archive (https://exoplanetarchive.ipac.caltech.edu/index.html) plus 16 planets that are not 
included in the archive:
KMT-2020-BLG-0414Lb \citep{Zang2021a},
OGLE-2019-BLG-1053Lb \citep{Zang2021b},
KMT-2019-BLG-0253Lb, KMT-2019-BLG-0953Lb, OGLE-2018-BLG-0977Lb, 
OGLE-2018-BLG-0506Lb, OGLE-2018-BLG-0516Lb, OGLE-2019-BLG-1492Lb \citep{Hwang2022},
KMT-2017-BLG-2509Lb,  OGLE-2017-BLG-1099Lb, OGLE-2019-BLG-0299Lb \citep{Han2021},
OGLE-2016-BLG-1093Lb \citep{Shin2022},
KMT-2018-BLG-1988Lb \citep{Han2022b},
KMT-2021-BLG-0912Lb \citep{Han2022a}, 
OGLE-2019-BLG-0468Lb, and OGLE-2019-BLG-0468Lc \citep{Han2022c}.},
80 (59\%) planets were found with the use of the data from the KMTNet survey.  This can be seen 
in Figure~\ref{fig:one}, in which we present the histogram of microlensing planets as a function 
of the planet-to-host mass ratio $q$.  The histogram also shows that the high-cadence surveys 
contribute to the detections of planets with very low mass ratios, especially those with $q<10^{-4}$.

In this paper, we report four planetary systems with low planet-to-host mass ratios detected 
from the microlensing surveys. Despite the short durations, ranging from a few hours to a couple 
of days, the planetary signals were clearly detected by the combined data of the lensing 
surveys. It is found that all of the discovered planets have sub-Jovian masses lying in the range 
of [0.05--0.38]~$M_{\rm J}$, illustrating the importance of the high-cadence surveys in detecting 
low-mass planets.

We present the analysis of the planetary lensing events according to the following organization. 
In Sect.~\ref{sec:two}, we describe the observations of the individual lensing events, instrument 
used for observations, and the process of the data reduction. In Sect.~\ref{sec:three}, we explain 
the procedure of analysis that is applied to each of the individual events. In the subsequent 
subsections, we describe the detailed features of the planetary signals and present the results 
of the analyses conducted for the individual lensing events.  In Sect.~\ref{sec:four}, we specify 
the types of the source stars and estimate the angular Einstein radii of the lens systems. In 
Sect.~\ref{sec:five}, we estimate the physical parameters of the planetary systems by conducting 
Bayesian analyses. We summarize the results of the analyses and conclude in Sect.~\ref{sec:six}.

\section{Observations and data}\label{sec:two}

The four lensing events for which we present analyses are (1) OGLE-2017-BLG-1691 (KMT-2017-BLG-0752), 
(2) KMT-2021-BLG-0320, (3) KMT-2021-BLG-1303 (MOA-2021-BLG-182), and (4) KMT-2021-BLG-1554.
The source stars of all events lie toward the Galactic bulge field. In Table~\ref{table:one}, 
we list the equatorial and galactic coordinates, observation fields of the surveys, alert 
dates, and baseline magnitudes of the individual events.  The baseline magnitude $I_{\rm base}$ 
is approximately scaled to the OGLE-III photometry system.

The events were detected from the combined observations of the lensing surveys conducted by the 
KMTNet, OGLE, and MOA groups. The KMTNet survey utilizes three identical 1.6~m telescopes that are
distributed in three countries of the Southern Hemisphere for the continuous coverage of lensing
events. The sites of the individual telescopes are the Siding Spring Observatory in Australia (KMTA), 
the Cerro Tololo InterAmerican Observatory in Chile  (KMTC), and the South African Astronomical 
Observatory in South Africa (KMTS).  Observations by the OGLE survey were conducted using the 
1.3~m telescope located at Las Campanas Observatory in Chile.  The MOA survey uses the 1.8~m 
telescope at the Mt.~John Observatory in New Zealand.  Observations were done mainly in the $I$ 
band for the KMTNet and OGLE surveys and in the customized MOA-$R$ band for the MOA survey. A subset 
of images were acquired in the $V$ band for the purpose of estimating source colors. The detailed 
procedure of the source color estimation will be discussed in Sect.~\ref{sec:four}.

\begin{table}[t]
\small
\caption{Error bar readjustment factors\label{table:two}}
\begin{tabular*}{\columnwidth}{@{\extracolsep{\fill}}llccc}
\hline\hline
\multicolumn{1}{c}{Event}                        &
\multicolumn{1}{c}{Data set}                     &
\multicolumn{1}{c}{$k$}                          &
\multicolumn{1}{c}{$\sigma_{\rm min}$ (mag)}     \\
\hline
 OGLE-2017-BLG-1691  & OGLE          & 1.010   & 0.010  \\
                     & KMTA          & 0.866   & 0.040  \\
                     & KMTC          & 1.170   & 0.010  \\
                     & KMTS          & 1.131   & 0.020  \\
\hline
 KMT-2021-BLG-0320   & KMTA (BLG01)  & 1.127   & 0.040  \\
                     & KMTA (BLG41)  & 1.389   & 0.025  \\
                     & KMTC (BLG01)  & 1.091   & 0.020  \\
                     & KMTC (BLG41)  & 1.175   & 0.020  \\
                     & KMTS (BLG01)  & 1.356   & 0.010  \\
                     & KMTS (BLG41)  & 1.319   & 0.010  \\
\hline
 KMT-2021-BLG-1303   & KMTA          & 1.304   & 0.020  \\
                     & KMTC          & 1.237   & 0.010  \\
                     & KMTS          & 1.364   & 0.010  \\
                     & MOA           & 1.034   & 0.020  \\
\hline
 KMT-2021-BLG-1554   & KMTA (BLG01)  & 1.371   & 0.020  \\
                     & KMTA (BLG41)  & 1.446   & 0.020  \\
                     & KMTC (BLG01)  & 1.418   & 0.020  \\
                     & KMTC (BLG41)  & 1.367   & 0.020  \\
                     & KMTS (BLG41)  & 1.364   & 0.020  \\
\hline                                             
\end{tabular*}
\end{table}

Reduction and photometry of the data were done using the pipelines of the individual survey 
groups: \citet{Albrow2009} for the KMTNet survey, \citet{Wozniak2000} for the OGLE survey, and 
\citet{Bond2001} for the MOA survey. All these pipelines apply the difference image analysis 
algorithm \citep{Tomaney1996, Alard1998} developed for the optimal photometry of stars lying in 
very dense star fields.  For a subset of the KMTC data set, we carried out additional photometry 
utilizing the pyDIA code \citep{Albrow2017} for the specification of the source colors.  Following 
the routine described in \citet{Yee2012}, we rescale the error bars of data by $\sigma=k
(\sigma_{\rm min}^2+\sigma_0^2)^{1/2}$, where $\sigma_0$ represents the error estimated from 
the photometry pipeline, $\sigma_{\rm min}$ is a factor used to make the data consistent with 
the scatter of data, and the factor $k$ is used to make $\chi^2$ per degree of freedom for 
each data set become unity.  In Table~\ref{table:two}, we list the factors $k$ and 
$\sigma_{\rm min}$ of the individual data sets.

\section{Analyses}\label{sec:three}

The light curves of the analyzed events share a common characteristic that short-lived anomalies
appear on the otherwise smooth and symmetric form of a single-lens single-source (1L1S) event.
Such anomalies in lensing light curves can be produced by two channels, in which the first
channel is a perturbation induced by a low-mass companion, such as a planet, to the lens 
\citep{Gould1992b}, and the second channel is an anomaly caused by a faint companion to the 
source \citep{Gaudi1998}. Hereafter we denote events with a binary lens and a binary source as 
2L1S and 1L2S events, respectively.  We examine the origins of the anomalies by modeling the 
light curve of the individual events under these 2L1S and 1L2S interpretations.

Lensing light curves are described by the combination of various lensing parameters. For 
a 1L1S event, the light curve is described by three parameters of $(t_0, u_0, t_{\rm E})$, 
which denote the time of the closest approach of the source to the lens, the separation 
(scaled to the angular Einstein radius $\thetae$) between the source and lens at that time, 
and the time scale of the event, respectively. The event time scale is defined as the time 
required for a source to transit $\thetae$.

In addition to these basic parameters, describing the light curve of a 2L1S event requires 
one to include extra parameters of $(s, q, \alpha)$, which represent the separation (scaled 
to $\thetae$) and mass ratio between the binary lens components, and the angle between the 
binary-lens axis and the direction of the source motion (source trajectory angle), respectively. 
Because planet-induced anomalies are usually generated by the crossings over or close approach 
of the source to the planet-induced caustics, an additional parameter of the normalized source 
radius $\rho$, which is defined as the ratio of the angular radius of the source 
$\theta_*$ to $\thetae$, is needed to describe the deformation of the anomaly by finite-source 
effects \citep{Bennett1996}.  In computing finite-source magnifications, we consider the 
limb-darkening variation of the source.

One also needs extra parameters to describe the lensing light curve of a 1L2S event. These 
parameters are $(t_{0,2}, u_{0,2}, q_F)$, which represent the closest approach time and 
separation between the lens and the second source, and the flux ratio between the binary 
source stars, respectively. In the 1L2S model, we designate the lensing parameters related 
to the primary source, $S_1$, as $(t_{0,1}, u_{0,1})$ to distinguish them from those related 
to the source companion, $S_2$.  In order to consider finite-source effects occurring when 
the lens passes over either of the source stars, we add two additional parameters of the 
normalized source radii $\rho_1$ and $\rho_2$ for the lens transits over $S_1$ and $S_2$, 
respectively \citep{Dominik2019}.

In modeling 1L1S and 1L2S light curves,  for which the lensing magnification is smooth with 
the variation of the lensing parameters, we search for the solution of the lensing parameters 
via a downhill approach by minimizing $\chi^2$ using the Markow Chain Monte Carlo (MCMC) 
algorithm. In modeling 2L1S light curves, for which the lensing magnification variation is 
discontinuous due to the formation of caustics and there may exist multiple local solutions 
caused by various types of degeneracy, we model the light curves in two steps. In the first step, 
we conduct grid searches for the binary parameters $s$ and $q$, construct a $\chi^2$ map on 
the $s$--$q$ parameter plane, and then identify local solutions appearing on the $\chi^2$ map. 
In the second step, we polish the individual local solutions using the downhill approach. Below 
we present the details of the modeling conducted for the individual lensing events.

\begin{figure}[t]
\includegraphics[width=\columnwidth]{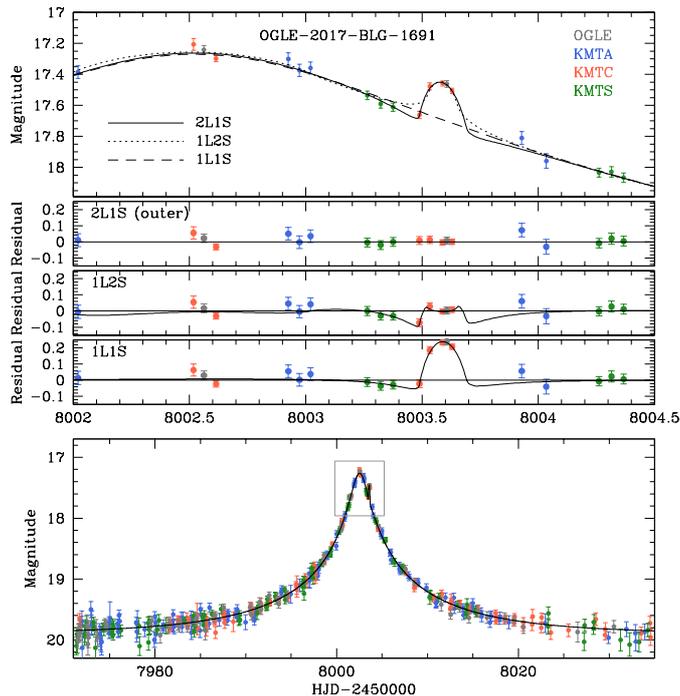}
\caption{
Lensing light curve of OGLE-2017-BLG-1691. The bottom panel shows the whole view, and the 
top panel displays the enlargement of the region around the anomaly (the region enclosed by 
a box in the bottom panel). In the top panel, curves of the three tested models, 2L1S (outer), 
1L2S, and 1L1S models, are drawn over the data points, and the residuals from the individual 
models are presented in the three middle panels. The curve drawn in the bottom panel is the 
best-fit model (outer 2L1S model). Colors of data points are set to match those of the 
telescopes used for observations marked in the legend. The curves drawn in the 1L2S and 1L1S 
residual panels represent the differences from the 2L1S model.
}
\label{fig:two}
\end{figure}

\subsection{OGLE-2017-BLG-1691 (KMT-2017-BLG-0752)}\label{sec:three-one}

The lensing event OGLE-2017-BLG-1691 occurred during the 2017 bulge season. It was first found
by the OGLE survey, and later confirmed by the KMTNet survey from the post-season analyses of
the data collected during the season.  The KMTNet group designated the event as KMT-2017-BLG-0752.
Hereafter, we use the nomenclatures of events by the ID references of the surveys who first found 
the events  in accordance with the convention of the microlensing community.  The baseline magnitude 
of the event was $I_{\rm base}=19.900\pm 0.004$.

The lensing light curve of OGLE-2017-BLG-1691 is shown in Figure~\ref{fig:two}, in which the 
bottom panel shows the whole view and the top panel displays the enlargement of the region 
around the anomaly. The event was alerted by the OGLE survey on 2017 September 6, 
${\rm HJD}^\prime\equiv {\rm HJD}-2450000 \sim 8002.5$, which approximately corresponds to the 
time near the peak.  An anomaly occurred about one day after the peak, but it was not noticed 
during the progress of the event because it was covered by just a single OGLE data point, and 
the KMTNet data were not released during the lensing magnification.  The existence of the anomaly 
was identified five years after the event, shortly after the re-reduction of all 2017 KMT light 
curves, from a project of reinvestigating the previous KMTNet data conducted to find unnoticed 
planetary signals \citep{Han2022b}. The top panel of Figure~\ref{fig:two} shows that the anomaly, 
which lasted for $\sim 5$~hours centered at ${\rm HJD}^\prime \sim 8003.6$, was additionally 
covered by three KMTC data points, and this confirmed that the single anomalous OGLE point was real.

\begin{figure}[t]
\includegraphics[width=\columnwidth]{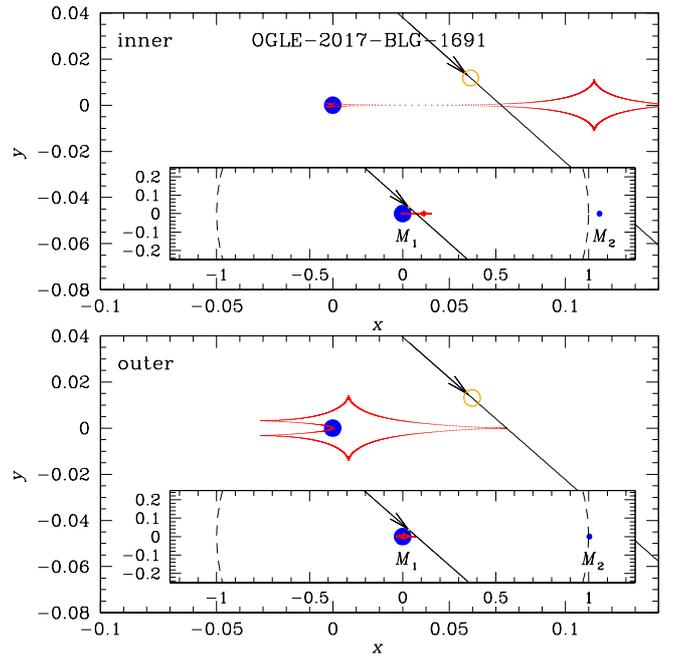}
\caption{
Lens system configurations of OGLE-2017-BLG-1691: upper panel (inner solution) and lower 
panel (outer solution).  The inset in each panel shows the whole view of the lens system, 
and the main panel presents the zoom-in view of the caustics.  The red cuspy figure is the 
caustic and the line with an arrow is the source trajectory.  The orange circle on the 
trajectory is drawn to represent the source scale with respect to the caustic.  The blue 
solid dots marked by $M_1$ and $M_2$ indicate the positions of the host and planet, 
respectively.  The dashed circle with unit radius in the inset represents the Einstein ring.
}
\label{fig:three}
\end{figure}

\begin{table*}[t]
\small
\caption{Lensing parameters of OGLE-2017-BLG-1691\label{table:three}}
\begin{tabular}{llll}
\hline\hline
\multicolumn{1}{c}{Parameter}           &
\multicolumn{1}{c}{2L1S (Inner)}        &
\multicolumn{1}{c}{2L1S (Outer)}        & 
\multicolumn{1}{c}{1L2S}                \\ 
\hline
$\chi^2$                  &   866.9                    &   866.5                    &   880.4                   \\
$t_{0}$ (HJD$^\prime$)    &   $8002.540 \pm  0.010  $  &  $8002.531 \pm 0.010  $    &  $8002.499 \pm 0.012  $   \\ 
$u_{0}$ ($10^{-2}$)       &   $4.83 \pm  0.27       $  &  $4.95 \pm 0.26       $    &  $6.55 \pm 0.49       $   \\
$\te$ (days)              &   $19.42 \pm 0.83       $  &  $18.94 \pm 0.82      $    &  $18.77 \pm 0.83      $   \\ 
$s$                       &   $1.058 \pm 0.011      $  &  $1.003 \pm 0.014     $    &  --                       \\
$q$ ($10^{-4}$)           &   $0.708 \pm 0.23       $  &  $0.97 \pm 0.34       $    &  --                       \\
$\alpha$ (rad)            &   $3.873 \pm  0.012     $  &  $3.865 \pm 0.010     $    &  --                       \\
$\rho$ ($10^{-3}$)        &   $3.40 \pm 0.47        $  &  $3.54 \pm 0.52       $    &  --                       \\ 
$t_{0,2}$ (HJD$^\prime$)  &   --                       &  --                        &  $8003.587 \pm 0.015  $   \\ 
$u_{0,2}$ ($10^{-2}$)     &   --                       &  --                        &  $0.029 \pm 0.124     $   \\
$\rho_2$ ($10^{-3}$)      &   --                       &  --                        &  $3.39 \pm 0.93       $   \\
$q_F$ ($10^{-3}$)         &   --                       &  --                        &  $6.11 \pm 1.85       $   \\ 
\hline
\end{tabular}
\tablefoot{ ${\rm HJD}^\prime = {\rm HJD}- 2450000$.  }
\end{table*}

We model the light curve of the event under the 1L1S, 2L1S and 1L2S interpretations.  The 
residuals of the individual tested models are compared in Figure~\ref{fig:two}.  It is found 
that the 2L1S model best describes the anomaly, being favored over the 1L1S and 1L2S models 
by $\Delta\chi^2=456.1$ and 13.9, respectively.  From the comparison of the residuals, it is 
found that the 2L1S model is confirmed not only by the 4 data points (3 KMTC plus 1 OGLE points) 
with positive deviations near the peak of the bump but also by the 3 additional points (2 
KMTS and 1 KMTC points) before the major bump with slightly negative deviations.

We find two degenerate 2L1S solutions with binary parameters of $(s, q) \sim (1.058, 0.71 
\times 10^{-4})$ and $\sim (1.003, 0.97 \times 10^{-4})$, which we designate as ``inner'' 
and ``outer'' solutions, respectively, for the reason to be mentioned below.  We list the 
full lensing parameters of the two 2L1S solutions in Table~\ref{table:three},  
together with the parameters of the 1L2S model.  The degeneracy is very 
severe and the outer solution is favored only by $\Delta\chi^2=0.4$.  The lensing parameters 
of the solutions indicate that the anomaly was produced by a very small mass-ratio planetary 
companion to the lens lying close to the Einstein ring of the primary regardless of the solutions.  
From the fact that both solutions have separations greater than unity and do not follow the 
relation of $s_{\rm inner}\times s_{\rm outer}\simeq 1$, the degeneracy is different from the 
``close-wide'' degeneracy \citep{Griest1998, Dominik1999, An2005}, which arises due to the 
similarity between the central caustics of planetary lens systems with planet separations $s$ 
and $1/s$.  Instead, the planet separations of the two degenerate solutions follow the relation
\begin{equation}
\sqrt{s_{\rm in}\times s_{\rm out}}=s^\dagger;\qquad
s^\dagger = {\sqrt{u_{\rm anom}^2+4}+u_{\rm anom} \over 2},
\label{eq1}
\end{equation}
where $u_{\rm anom}^2=\tau_{\rm anom}^2+u_0^2$, $\tau_{\rm anom}=(t_{\rm anom}-t_0)/\te$, and 
$t_{\rm anom}$ indicates the time of the planetary anomaly \citep{Hwang2022, Zhang2022, Ryu2022}.  
With the values $t_0\simeq 8002.5$, $u_0 \simeq 4.85\times 10^{-2}$, $\te\simeq 19$~day and 
$t_{\rm anom}\simeq 8003.6$, one finds that $s^\dagger\simeq 1.04$, which matches well 
$\sqrt{s_{\rm in}\times s_{\rm out}}\simeq 1.03$.  This indicates that the similarity between 
the light curves of the two solutions is caused by the degeneracy identified by \citet{Yee2021}, 
who first mentioned the continuous transition between the ``close-wide'' and ``inner-outer'' 
\citep{Gaudi1997} degeneracies.  Hereafter, we refer to this degeneracy as ``offset degeneracy'' 
following \citet{Zhang2022}.

Figure~\ref{fig:three} shows the lens systems configurations, in which the source trajectory 
(line with an arrow) with respect to the caustic and positions of the lens components (blue 
solid dots marked by $M_1$ and $M_2$) is presented: inner solution in the upper panel and 
outer solution in the lower panel.  The source passed the inner side (with respect to $M_1$) 
of the caustic according to the inner solution, while the source passes the outer side according 
to the outer solution.  For both solutions, the source crossed the cusp of the caustic, and thus 
the anomaly is affected by finite-source effects during the caustic crossing, allowing us to 
precisely measure the normalized source radius.  For each solution, the source size relative 
to the caustic is represented by an orange circle marked on the source trajectory. As will be 
discussed in Sect.~\ref{sec:four}, the measurement of $\rho$ is important to estimate the 
lensing observable of the angular Einstein radius $\thetae$, which can be used to constrain 
the physical lens parameters.  However, the microlens parallax vector, $\pie=(\pi_{\rm rel}/
\thetae)/(\muvec/\mu)$, which is another observable constraining the physical lens parameters, 
cannot be securely measured because the event time scale, $\te\sim 19$~days, is not long enough 
to produce detectable deviations induced by the orbital motion of Earth around the Sun
\citep{Gould1992a}. Here $\muvec$ represents the vector of the relative 
lens-source proper motion.

\subsection{KMT-2021-BLG-0320}\label{sec:three-two}

The lensing event KMT-2021-BLG-0320 was found on 2021 April 9 (${\rm HJD}^\prime \sim 9313$), 
when the event had not yet reached its peak, with the employment of the KMTNet AlertFinder 
system \citep{Kim2018}, which began full operation since the 2019 season.  Two days after the 
detection, the event reached its peak with a magnification of $A_{\rm max}\sim 170$, and then 
gradually declined until it reached its baseline of $I_{\rm base}=20.08$.  The source of the 
event lies in the two KMTNet prime fields of BLG01 and BLG41, toward which observations were 
conducted with a 30~min cadence for each field, and thus with a 15~min combined cadence. The 
areas covered by the two fields overlap except for about 15\% of the area of each field filling 
the gaps between the chips of the camera.  Because the data were taken from two fields of three 
telescopes, there are 6 data sets: KMTA (BLG01), KMTA (BLG41), KMTC (BLG01), KMT (BLG41), KMTS 
(BLG01), and KMTS (BLG41).

\begin{figure}[t]
\includegraphics[width=\columnwidth]{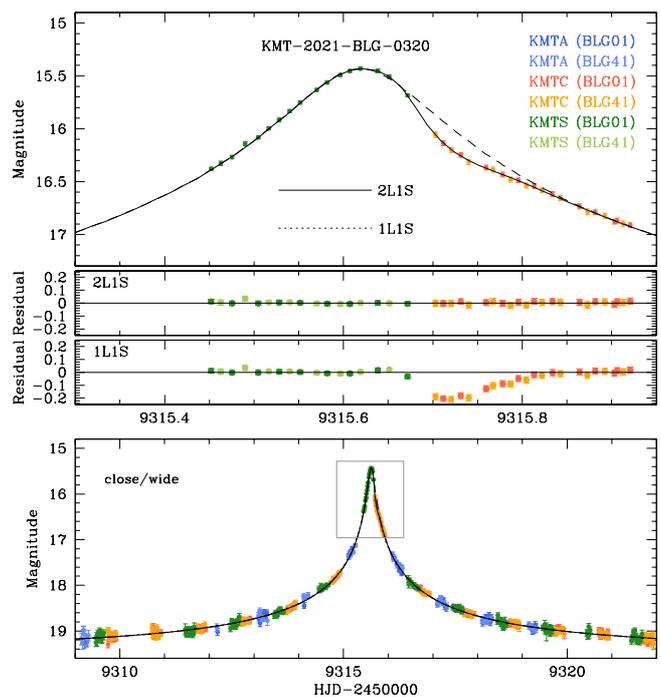}
\caption{
Lensing light curve of KMT-2021-BLG-0320. Notations are same as those in Fig.~\ref{fig:two}.
}
\label{fig:four}
\end{figure}

\begin{table}[t]
\small
\caption{Lensing parameters of KMT-2021-BLG-0320\label{table:four}}
\begin{tabular*}{\columnwidth}{@{\extracolsep{\fill}}llll}
\hline\hline
\multicolumn{1}{c}{Parameter}    &
\multicolumn{1}{c}{Close}        &
\multicolumn{1}{c}{Wide}         \\
\hline
$\chi^2$                &   4075.5                  &   4075.5                       \\
$t_{0}$ (HJD$^\prime$)  &  $9315.615 \pm 0.001   $  &  $9315.615 \pm 0.001   $       \\ 
$u_{0}$ ($10^{-3}$)     &  $5.86 \pm 0.21        $  &  $5.93 \pm 0.21        $       \\
$\te$ (days)            &  $13.12 \pm 0.39       $  &  $12.97 \pm 0.40       $       \\ 
$s$                     &  $0.771 \pm 0.014      $  &  $1.274 \pm 0.024      $       \\
$q$ ($10^{-4}$)         &  $3.02 \pm 0.42        $  &  $2.89 \pm 0.44        $       \\
$\alpha$ (rad)          &  $0.604 \pm 0.014      $  &  $0.602 \pm 0.014      $       \\
$\rho$ ($10^{-3}$)      &  --                       &  --                            \\ 
\hline
\end{tabular*}
\end{table}

In Figure~\ref{fig:four}, we present the light curve constructed from the combination of the 
6 KMTNet data sets.  Although it would be difficult to notice the anomaly from a glimpse, we 
inspected the light curve because the event reached a very high magnification at the peak, near 
which the light curve is susceptible to perturbations induced by a planet \citep{Griest1998}.  
From this inspection, it was found that the light curve exhibited an anomaly that lasted for 
about 4 hours with a negative deviation with respect to a 1L1S model.  See the enlarged view 
around the peak region of the light curve presented in the top panel of Figure~\ref{fig:four}.  
It is known that a planetary companion to a lens can induce anomalies with both positive and 
negative deviations, while a faint companion to a source can induce anomalies with only positive 
deviations \citep{Gaudi1998}. We, therefore, modeled the light curve under the 2L1S interpretation.

It is found that the anomaly is well described by a 2L1S model, in which the mass ratio between 
the binary lens components is very low. We found two sets of solutions with $(s, q)_{\rm close}
\sim (0.77, 3.0\times 10^{-4})$ and $(s, q)_{\rm wide} \sim (1.27, 2.9\times 10^{-4})$.  We 
designate the individual solutions as ``close'' and ``wide'' solutions, because the former 
solution has $s<1.0$ and the latter has $s>1.0$. In Table~\ref{table:four}, we list the full 
lensing parameters for the two sets of solutions. The fits of the two solutions are nearly 
identical with $\chi^2$ values that are equal to the first digit after the decimal point, 
indicating that the degeneracy between the two solutions is very severe.  The fact that the 
binary separations of the two solutions follow the relation of $s_{\rm close}\times s_{\rm wide} 
\simeq 1.0$ indicates that the similarity between the solution stems from the close-wide degeneracy.  
It is known that the relation between the planet separations of the solutions under the offset 
degeneracy in Equation~(\ref{eq1}) applies to more general cases, including resonant case, than 
the $s_{\rm close}\times s_{\rm wide} \simeq 1.0$ relation of the close-wide degeneracy, and 
we confirm this.  In addition to the relation between $s_{\rm in}$ and $s_{\rm out}$, 
\citet{Hwang2022} provided analytic formulas for the heuristic estimation of the source 
trajectory angle and the mass ratio; 
\begin{equation}
\alpha = \tan^{-1}\left( {u_0 \over \tau_{\rm anom}}\right);\qquad
q = \left( {\Delta t_{\rm anom}\over 4\te} \right)  \left( {s\over |u_0|}\right) 
\left\vert \sin^3 \alpha\right\vert,
\label{eq2}
\end{equation}
where $\tau_{\rm anom}$ denotes the duration of the planet-induced anomaly.  We confirm that 
the heuristically estimated values of $\alpha$ and $q$ match well those estimated from the 
modeling.  It was found that the normalized source radius could not be constrained, not even 
the upper limit, and thus the line for $\rho$ in Table~\ref{table:four} is left blank.  

\begin{figure}[t]
\includegraphics[width=\columnwidth]{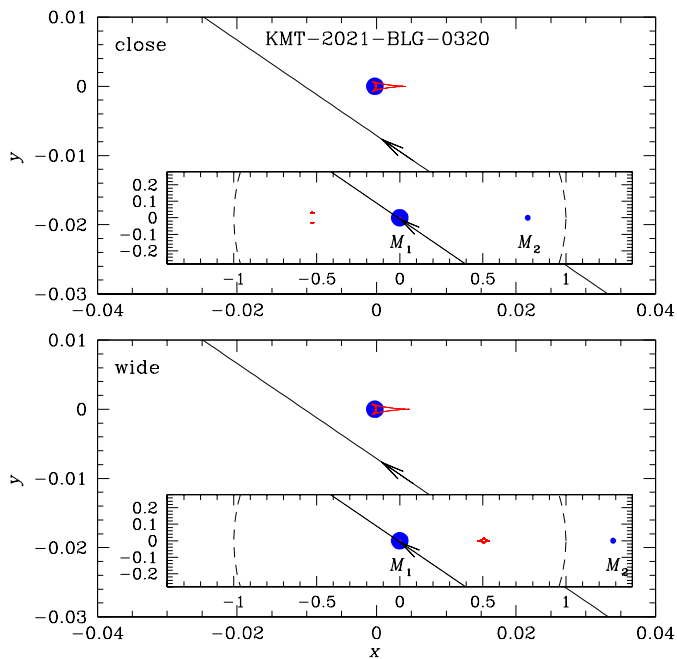}
\caption{
Lens system configurations of KMT-2021-BLG-0320.  The upper and lower panels show the configurations 
of the close and wide solutions, respectively.
}
\label{fig:five}
\end{figure}

Figure~\ref{fig:five} shows the lensing configuration of KMT-2021-BLG-0320 for the close (upper 
panel) and wide (lower panel) solutions. It shows that the source passed the back-end region of 
the tiny central caustic induced by a planet, and this generated the negative deviation of the 
observed anomaly. The central caustics of the close and wide solutions are very similar, resulting 
in nearly identical deviations.  Because the source passed well outside the caustic, finite-source 
effects could not be detected, and thus we do not mark an orange circle representing the source 
size.  Microlens-parallax effects could not be securely detected because the event time scale, 
$\te \sim 13$~days, is much shorter than the orbital period of Earth, that is, 1~year.

\begin{figure}[t]
\includegraphics[width=\columnwidth]{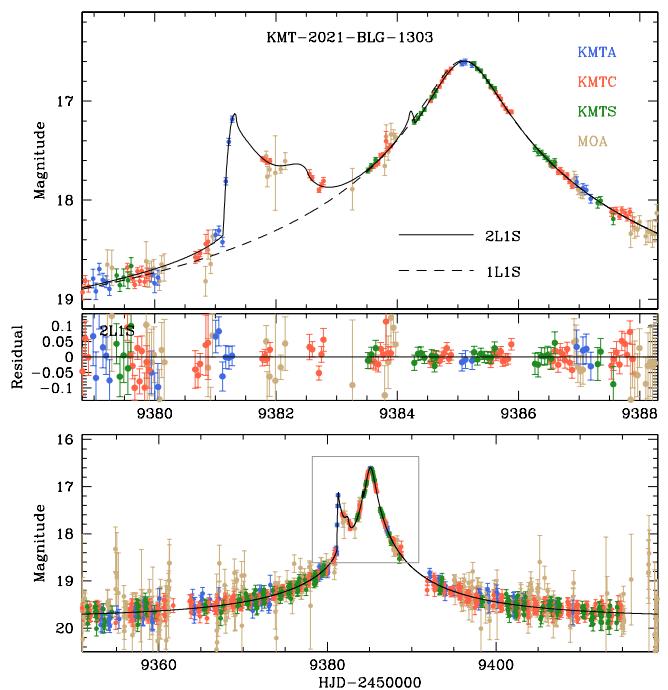}
\caption{
Lensing light curve of KMT-2021-BLG-1303. Notations are same as those in Fig.~\ref{fig:two}.
}
\label{fig:six}
\end{figure}

\subsection{KMT-2021-BLG-1303 (MOA-2021-BLG-182)}\label{sec:three-three}

The event KMT-2021-BLG-1303 was first found on 2021 June 14 (${\rm HJD}^\prime \sim 9379$) by the 
KMTNet survey, and later identified by the MOA survey on 2021 June 17 (${\rm HJD}^\prime \sim 9382$). 
The MOA survey designated the event as MOA-2021-BLG-182. A day after the first discovery, the event
displayed an anomaly that lasted for about 2~days. After the anomaly, the event reached peak at
${\rm HJD}^\prime \sim 9385.1$ with a magnification of $A_{\rm max}\sim 46$, and then gradually 
declined to the baseline of $I_{\rm base}=19.67$.

The light curve of KMT-2021-BLG-1303 constructed with the combined KMTNet and MOA data is shown in 
Figure~\ref{fig:six}. Compared to the previous two events, the light curve of this event displays 
a very obvious anomaly with a maximum deviation of $\Delta I\sim 1.4$~mag from a 1L1S model. The 
rapid brightening at ${\rm HJD}^\prime \sim 9381.0$ indicates that the event experienced a caustic 
crossing. Because caustics form due to the multiplicity of lens components, we rule out the 1L2S 
origin of the anomaly and test only 
the 2L1S interpretation.

\begin{table}[t]
\small
\caption{Lensing parameters of KMT-2021-BLG-1303\label{table:five}}
\begin{tabular*}{\columnwidth}{@{\extracolsep{\fill}}llll}
\hline\hline
\multicolumn{1}{c}{Parameter}    &
\multicolumn{1}{c}{Value }        \\
\hline
$\chi^2$                &   1654.9                    \\
$t_{0}$ (HJD$^\prime$)  &  $9385.091 \pm 0.003  $     \\ 
$u_{0}$ ($10^{-2}$)     &  $2.17 \pm 0.06       $     \\
$\te$ (days)            &  $25.21 \pm 0.68      $     \\ 
$s$                     &  $1.029 \pm 0.001     $     \\
$q$ ($10^{-4}$)         &  $6.42 \pm 0.45       $     \\
$\alpha$ (rad)          &  $6.138 \pm 0.001     $     \\
$\rho$ ($10^{-3}$)      &  $0.95 \pm 0.08       $     \\ 
\hline
\end{tabular*}
\end{table}

We find that the anomaly is well explained by a 2L1S model with a planetary companion
lying close to the Einstein ring of the host.  We find a unique solution without any
degeneracy, and the binary parameters of the solution are $(s, q)\sim (1.029, 6.4\times 
10^{-4})$.  Here again the planet-to-host mass ratio is of the order of $10^{-4}$ like 
the two previous events.  We double checked the uniqueness of the solution by thoroughly 
inspecting the region around the $s$ and $q$ parameters predicted by the relations in 
Equations~(\ref{eq1}) and (\ref{eq2}), and confirmed that there is only a single solution.  
The full lensing parameters of the event are listed in Table~\ref{table:five}.

\begin{figure}[t]
\includegraphics[width=\columnwidth]{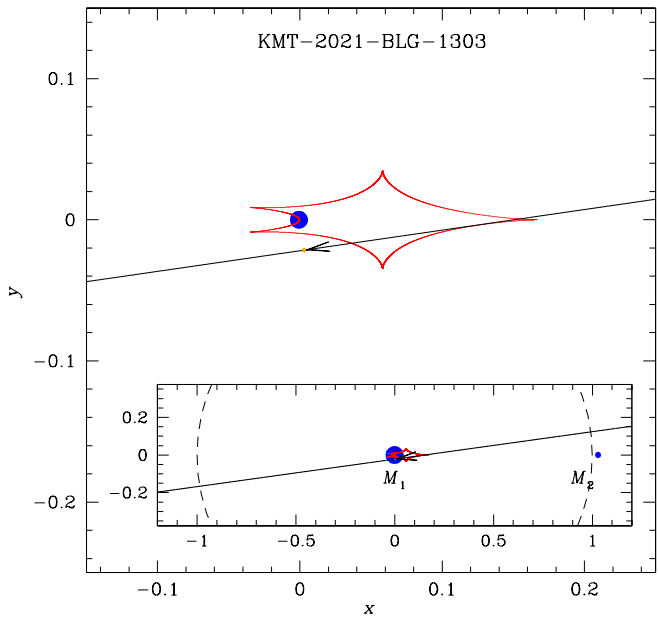}
\caption{
Lens system configuration of KMT-2021-BLG-1303. Notations are same as those in Fig.~\ref{fig:three}.
}
\label{fig:seven}
\end{figure}

\begin{table*}[t]
\small
\caption{Lensing parameters of KMT-2021-BLG-1554\label{table:six}}
\begin{tabular}{llll}
\hline\hline
\multicolumn{1}{c}{Parameter}         &
\multicolumn{1}{c}{2L1S (Close)}      &
\multicolumn{1}{c}{2L1S (Wide)}       &
\multicolumn{1}{c}{1L2S}              \\
\hline
$\chi^2$                  &   3486.0               &   3485.7                    &   3496.1                  \\
$t_{0}$ (HJD$^\prime$)    &  $9394.737 \pm 0.010$  &  $9394.742 \pm 0.010 $      &  $9394.719 \pm 0.019  $   \\ 
$u_{0}$                   &  $0.051 \pm 0.009   $  &  $0.051 \pm 0.011    $      &  $0.003 \pm 0.054     $   \\
$\te$ (days)              &  $5.04 \pm 0.78     $  &  $5.15 \pm 0.72      $      &  $6.55 \pm 1.11       $   \\ 
$s$                       &  $0.888 \pm 0.039   $  &  $1.189 \pm 0.051    $      &  --                       \\
$q$ ($10^{-4}$)           &  $14.14 \pm 4.48    $  &  $14.94 \pm 5.41     $      &  --                       \\
$\alpha$ (rad)            &  $4.705 \pm 0.038   $  &  $4.727 \pm 0.037    $      &  --                       \\
$\rho$ ($10^{-3}$)        &  $6.98 \pm 1.18     $  &  $7.00 \pm 1.28      $      &  $98.8 \pm 24.2       $   \\ 
$t_{0,2}$ (HJD$^\prime$)  &   --                   &  --                         &  $9394.740 \pm 0.002  $   \\ 
$u_{0,2}$                 &   --                   &  --                         &  $0.002 \pm 0.002     $   \\
$\rho_2$ ($10^{-3}$)      &   --                   &  --                         &  --                       \\
$q_F$                     &   --                   &  --                         &  $0.048 \pm 0.008     $   \\ 
\hline
\end{tabular}
\end{table*}

The lensing configuration of the event is presented in Figure~\ref{fig:seven}, which shows 
that the planet lies very close to the Einstein ring and induces a single six-sided resonant 
caustic near the primary of the lens. The source entered the caustic at ${\rm HJD}^\prime 
\sim 9381.0$, and this produced the sharp rise of the light curve at the corresponding time.  
According to the model, the source exited the caustic at ${\rm HJD}^\prime \sim 9384.2$, but 
the caustic-crossing feature at this epoch is not obvious in the lensing light curve due to 
the combination of the weak caustic and the poor coverage of the region. The pattern of the 
anomaly between the caustic entrance and exit deviates from a typical ``U''-shape pattern 
because the source passed along the fold of the caustic.  The light curve during the caustic 
entrance of the source was well resolved by the KMTA data, and thus the normalized source 
radius, $\rho=(0.95\pm 0.08)\times 10^{-3}$, is  tightly constrained.  However, it was difficult 
to constrain the microlens parallax because the time scale, $\te \sim 25$~days, is not long 
enough and the photometric precision of the faint source with $I\sim 21$ is not high enough 
to detect subtle deviations induced by microlens-parallax effects.

\begin{figure}[t]
\includegraphics[width=\columnwidth]{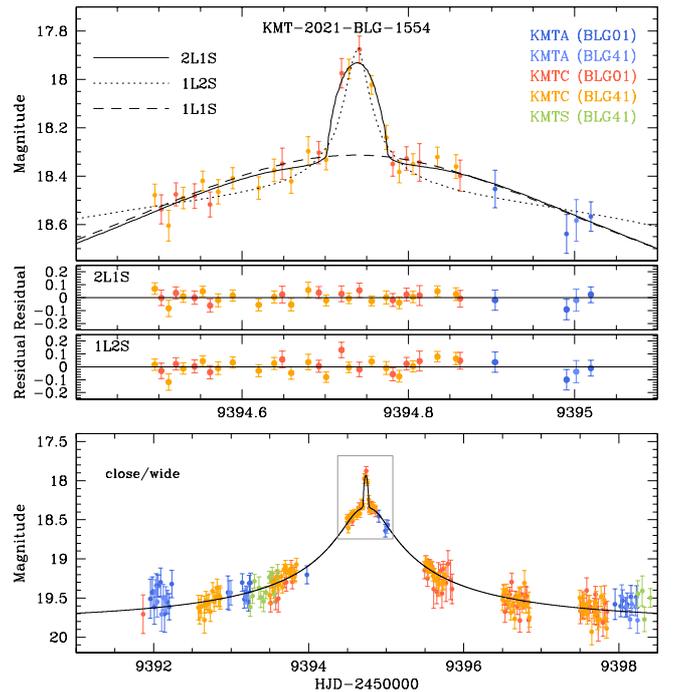}
\caption{
Lensing light curve of KMT-2021-BLG-1554. Notations are same as those in Fig.~\ref{fig:two}.
}
\label{fig:eight}
\end{figure}

\subsection{KMT-2021-BLG-1554}\label{sec:three-four}

This short event with a time scale of $\te\sim 5$~days and a faint baseline magnitude of 
$I_{\rm base}=22.39$ was detected on 2021 Jul~1 (${\rm HJD}^\prime \sim 9396$), about two 
days after the event peaked at $t_0\sim 9394.7$, by the KMTNet survey. The source lies in 
the prime fields of BLG01 and BLG41, and thus there are 6 data sets from the three KMTNet 
telescopes. Among these data sets, the data set of the BLG41 field acquired by the KMTS 
telescope was not used in the analysis due to its poor photometric quality caused by bad 
seeing during the lensing magnification.  Although the other data set from KMTS and those 
from KMTA were included in the analysis, the results are mainly derived from the KMTC data 
sets, which cover the peak region of the light curve.

The light curve constructed with the available KMTNet data sets is shown in Figure~\ref{fig:eight}. 
In the peak region, it exhibits an anomaly that lasted for about 3~hours from a 1L1S model with 
a positive deviation of $\Delta I\sim 0.5$~mag.  The anomaly appears both in the BLG01 
(two data points) and BLG41 (3 points) data sets obtained from KMTC observations, confirming 
that the signal is real.  Because a positive anomaly can be produced by a binary companion to 
either a lens or a source, we test both 2L1S and 1L2S models.

\begin{table*}[t]
\caption{De-reddened colors, magnitudes, and spectral types of source stars\label{table:seven}}
\begin{tabular}{llllll}
\hline\hline
\multicolumn{1}{c}{Event}       &
\multicolumn{1}{c}{$(V-I)_0$}   &
\multicolumn{1}{c}{$I_0$}       &
\multicolumn{1}{c}{Spectral type}        \\
\hline
 OGLE-2017-BLG-1691  &  $0.714 \pm 0.065$ &  $17.397 \pm   0.039$  &  G1 turnoff or subgiant   \\
 KMT-2021-BLG-0320   &  $0.700 \pm 0.040$ &  $18.843 \pm   0.002$  &  G1V  \\
 KMT-2021-BLG-1303   &  $0.849 \pm 0.015$ &  $19.809 \pm   0.003$  &  G9V  \\
 KMT-2021-BLG-1554   &  $0.873 \pm 0.061$ &  $18.649 \pm   0.008$  &  K1V  \\ 
\hline                                             
\end{tabular}
\end{table*}

\begin{figure}[t]
\includegraphics[width=\columnwidth]{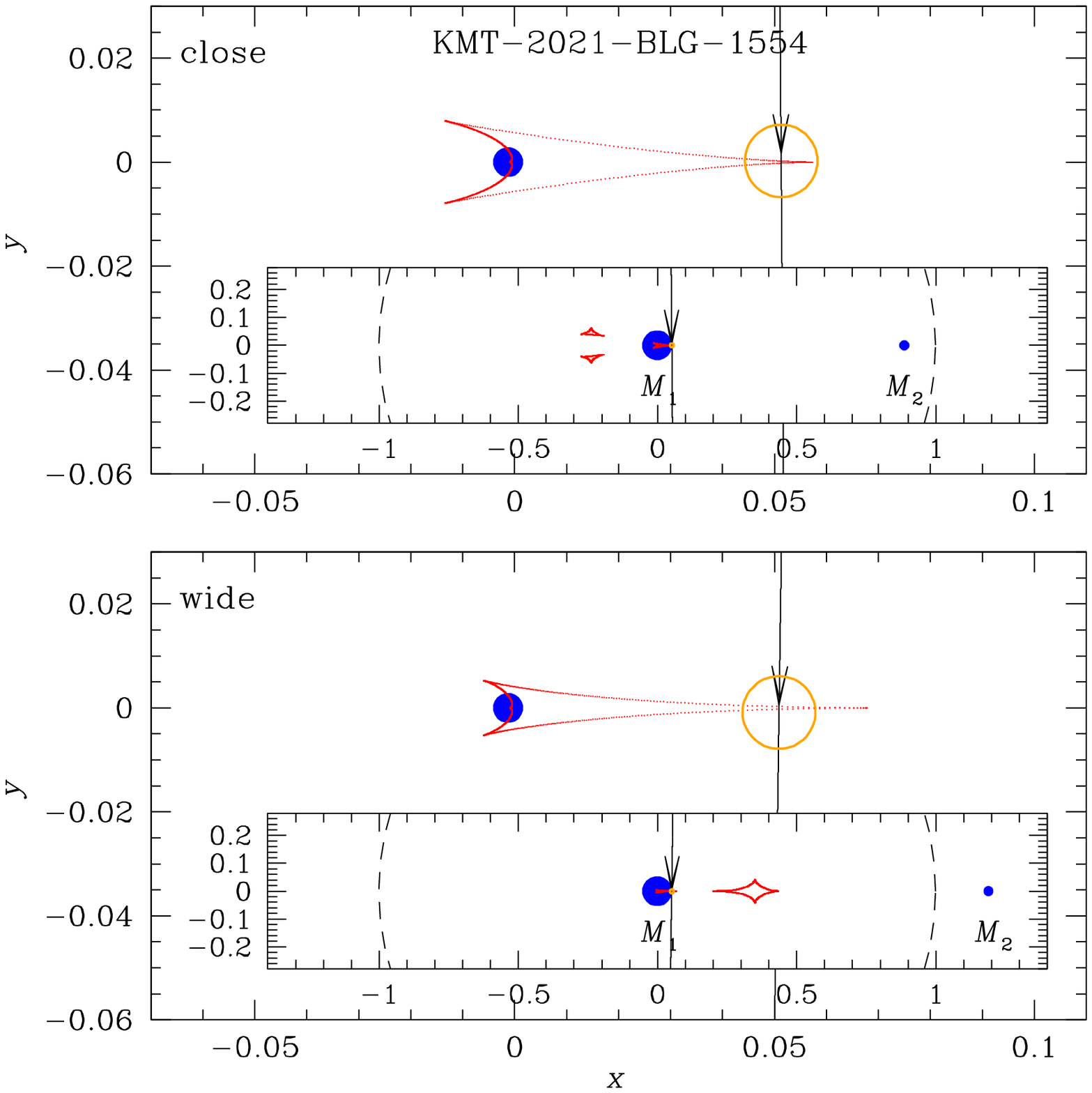}
\caption{
Lens system configuration of KMT-2021-BLG-1554. Notations are same as those in Fig.~\ref{fig:five}.
}
\label{fig:nine}
\end{figure}

\begin{table*}[t]
\caption{Angular source radii, Einstein radii, and relative lens-source proper motions \label{table:eight}}
\begin{tabular}{llllll}
\hline\hline
\multicolumn{1}{c}{Event}                  &
\multicolumn{1}{c}{$\theta_*$ ($\mu$as)}   &
\multicolumn{1}{c}{$\thetae$ (mas)}        &
\multicolumn{1}{c}{$\mu$ (mas yr$^{-1}$)}  \\
\hline
 OGLE-2017-BLG-1691  &  $1.043 \pm  0.100$ &  $0.30 \pm 0.05$  &  $5.68 \pm 1.01$  \\
 KMT-2021-BLG-0320   &  $0.528 \pm  0.042$ &  --               &  --               \\
 KMT-2021-BLG-1303   &  $0.401 \pm  0.029$ &  $0.42 \pm 0.05$  &  $6.13 \pm 0.70$  \\
 KMT-2021-BLG-1554   &  $0.703 \pm  0.065$ &  $0.10 \pm 0.02$  &  $7.30 \pm 1.50$  \\ 
\hline                                             
\end{tabular}
\end{table*}

Detailed analysis of the light curve indicates that the anomaly was produced by a planetary
companion to the lens.  We find two sets of solutions with planet parameters of $(s, q)_{\rm close}
\sim (0.89, 1.4\times 10^{-3})$ and $(s, q)_{\rm wide}\sim (1.12, 1.5\times 10^{-3})$ resulting 
from the close-wide degeneracy. The full lensing parameters of both solutions are provided in 
Table~\ref{table:six}.  It is found that the degeneracy between the two 2L1S solution is severe 
with $\Delta\chi^2=0.3$, and the 2L1S solutions are favored over the model under the 1L2S 
interpretation by $\Delta\chi^2\sim 11$.  The lensing parameters of the 1L2S model are listed 
in Table~\ref{table:six}.  We compare the residuals of the 2L1S and 1L2S models in the two middle 
panels of Figure~\ref{fig:eight}.  It turns out that the source crossed a caustic, and thus we 
are able to measure the normalized source radius, $\rho = (6.98\pm 1.18)\times 10^{-3}$.

The lensing configurations of the event according to the close and wide solutions are shown in the
upper and lower panels of Figure~\ref{fig:nine}, respectively.  The figure shows that the anomaly 
was produced by the crossing of the source over the sharp tip of the central caustic induced by a 
planet lying near the Einstein ring of the planet host.  The gap between the caustic entrance and 
exit is much smaller than the source size, and thus the individual caustic-crossing features do not 
show up in the anomaly and instead the anomaly appears as a single bump.  The incidence angle of the 
source on the binary axis is nearly $90^\circ$, and thus the features on the rising and falling sides 
of the anomaly are symmetric.

\section{Source stars and angular Einstein radii}\label{sec:four}

Among the four analyzed planetary events, the anomalies of the three events (KMT-2017-BLG-0752,
KMT-2021-BLG-1303, and KMT-2021-BLG-1554) were affected by finite-source effects, and thus the
normalized source radii can be measured. In this section, we estimate the angular Einstein 
radii for these events. Estimating $\thetae$ from a measured $\rho$ value requires one to specify 
the source type, from which the angular source radius $\theta_*$ is deduced and the Einstein 
radius is determined by
\begin{equation}
\thetae = {\theta_*\over \rho}.
\label{eq3}
\end{equation}
Although the normalized source radius and thus $\thetae$ cannot be measured for the event 
KMT-2021-BLG-0320, we specify the source type for the sake  of completeness.

The specification of the source type is done by estimating the color and brightness of the 
source.  We measure the $I$ and $V$-band magnitudes of the source from the regression of 
the data processed using the pyDIA photometry code \citep{Albrow2017} with the variation 
of the lensing magnification. We then place the source on the instrumental color-magnitude 
diagram (CMD) of stars lying around the source constructed using the same photometry code. 
Following the method of \citet{Yoo2004}, the instrumental source color and magnitude, 
$(V-I, I)$, are then calibrated using the centroid of red giant clump (RGC), for which its 
reddening and extinction-corrected (de-reddened) color and magnitude, $(V-I, I)_{\rm RGC,0}$, 
are known, on the CMD as a reference.

\begin{figure}[t]
\includegraphics[width=\columnwidth]{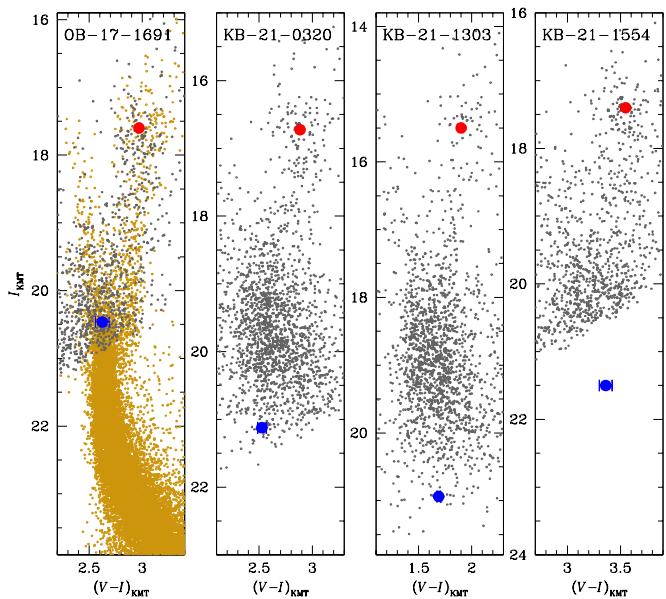}
\caption{
Source locations with respect to the centroids of the red giant clump (RGC) on the instrumental 
color-magnitude diagrams of stars lying around the source stars of the events OGLE-2017-BLG-1691,
KMT-2021-BLG-0320, KMT-2021-BLG-1303, and KMT-2021-BLG-1554.  For OGLE-2017-BLG-1691, the CMD is 
constructed by combining the two CMDs from KMTC (grey dots) and {\it HST} (brown dots) observations.  
For each event, the locations of the source and RGC centroid are marked by blue and red solid dots, 
respectively.
}
\label{fig:ten}
\end{figure}

Figure~\ref{fig:ten} shows the positions of the source stars (marked by blue dots) with 
respect to those of the RGC centroids (red dots) on the CMDs of the individual events.  For 
OGLE-2017-BLG-1691, the source color could not be securely measured not only because the 
$V$-band data sparsely covered the light curve during the lensing magnification but also 
because the photometry quality of these data is low, although the $I$-band magnitude was 
relatively well measured.  In this case, we estimated the source color as the median 
color of stars with the same $I$-band magnitude offset from the RGC centroid in the CMD constructed 
from the images of Baade's window taken from the observations using the {\it Hubble Space Telescope} 
\citep{Holtzman1998}.  From the offsets in color and magnitudes between the source and RGC centroid, 
$\Delta (V-I, I)$, together with the known de-reddened color and magnitude of the RGC centroid, 
$(V-I, I)_{\rm RGC,0}$ \citep{Bensby2013, Nataf2013}, we estimate the calibrated source color and 
magnitude as $(V-I, I)_0=(V-I, I)_{\rm RGC,0}+\Delta (V-I, I)$.  In Table~\ref{table:seven}, we list 
the source colors and magnitudes of the individual events estimated from this procedure along with 
the spectral types of the source stars.  It is found that the source of OGLE-2017-BLG-1691 is a 
turnoff star or a subgiant of a G spectral type, while the source stars of the other events 
are main-sequence stars with spectral types ranging from G to K.

For each event, the angular source radius and Einstein radius were estimated from the measured 
color and magnitude of the source.  For this, the measured $V-I$ color was converted into $V-K$ 
color using the color-color relation of \citet{Bessell1988}, $\theta_*$ value was interpolated 
from the ($V-K$)--$\theta_*$ relation of \citet{Kervella2004}, and then the angular Einstein 
radius was estimated using the relation in Equation~(\ref{eq3}).  With the measured Einstein 
radius together with the event time scale, the value of the relative lens-source proper motion 
was assessed as 
\begin{equation}
\mu = {\thetae\over \te}.
\label{eq4}
\end{equation}
In Table~\ref{table:eight}, we list the estimated values of $\theta_*$, $\thetae$, and $\mu$ 
for the individual events. The $\thetae$ and $\mu$ values for the event KMT-2021-BLG-0320 are 
left blank because finite-source effects in the lensing light curve were not detected and 
subsequently the values of $\rho$, $\thetae$, and $\mu$ could not be measured. We note that 
the angular Einstein radius of KMT-2021-BLG-1554, $\thetae\sim 0.10$~mas, is substantially 
smaller than those of the other events, and this together with its short time scale, 
$\te \sim 5$~days, suggests that the mass of the lens would be very low.  KMT-2021-BLG-1554 is 
the seventh shortest microlensing event with a bound planet. See Table~3 of \citet{Ryu2021}.

\begin{table*}[t]
\caption{Physical lens parameters\label{table:nine}}
\begin{tabular}{llllll}
\hline\hline
\multicolumn{1}{c}{Event}                           &
\multicolumn{1}{c}{$M_{\rm host}$ ($M_\odot$)}      &
\multicolumn{1}{c}{$M_{\rm planet}$ ($M_{\rm J}$)}  &
\multicolumn{1}{c}{$\dl$ (kpc)}                     &
\multicolumn{1}{c}{$a_\perp$ (AU)}                  \\
\hline
 OGLE-2017-BLG-1691  &  $0.45^{+0.36}_{-0.25}$  &  $0.046^{+0.037}_{-0.025}$    &  $7.29^{+1.01}_{-1.33}$  &  $2.41^{+0.34}_{-0.44}$ (inner)   \\  [0.7ex]
                     &  --                      &  --                           &  --                      &  $2.54^{+0.36}_{-0.46}$ (outer)   \\  [0.7ex]                            
 KMT-2021-BLG-0320   &  $0.32^{+0.39}_{-0.21}$  &  $0.10 ^{+0.13 }_{-0.07 }$    &  $6.95^{+1.07}_{-1.33}$  &  $1.54^{+0.24}_{-0.30}$ (close)   \\  [0.7ex]
                     &  --                      &  --                           &  --                      &  $2.55^{+0.39}_{-0.49}$ (wide)    \\  [0.7ex]                            
 KMT-2021-BLG-1303   &  $0.57^{+0.32}_{-0.29}$  &  $0.38 ^{+0.22 }_{-0.20 }$    &  $6.28^{+0.95}_{-1.43}$  &  $2.89^{+0.44}_{-0.66}$           \\  [0.7ex]
 KMT-2021-BLG-1554   &  $0.08^{+0.13}_{-0.04}$  &  $0.12 ^{+0.20}_{-0.07}$      &  $7.68^{+1.04}_{-1.10}$  &  $0.72^{+0.13}_{-0.14}$ (close)   \\  [0.7ex]
                     &   --                     &  --                           &  --                      &  $0.96^{+0.15}_{-0.16}$ (wide)    \\  [0.7ex]                    
\hline                                             
\end{tabular}
\end{table*}

\section{Physical parameters}\label{sec:five}

In this section, we estimate the physical parameters of the planetary systems.  For the unique 
constraint of the physical lens parameters, one must simultaneously measure the extra observables 
of $\thetae$ and $\pie$, from which the mass $M$ and distance to the lens, $\dl$, are determined as
\begin{equation}
M={\thetae \over \kappa \pie};\qquad
\dl = { {\rm AU} \over  \pie\thetae + \pi_{\rm S}}.
\label{eq5}
\end{equation}
Here $\kappa=4G/(c^2{\rm AU})$, $\pi_{\rm S}={\rm AU}/D_{\rm S}$, and $D_{\rm S}$ denotes the 
distance to the source \citep{Gould1992a, Gould2000}.  The values of the microlens parallax 
could not be measured for any of the events, and thus the physical parameters cannot be 
unambiguously determined from the relation in Equation~(\ref{eq5}).  However, one can still 
constrain $M$ and $\dl$ using the other measured observables of $\te$ and $\thetae$, which are 
related to the physical lens parameters as
\begin{equation}
\te = {\thetae \over \mu};\qquad
\thetae = (\kappa M \pi_{\rm rel})^{1/2},
\label{eq6}
\end{equation}
where $\pi_{\rm rel}={\rm AU}(D_{\rm L}^{-1}-D_{\rm S}^{-1})$ denotes the relative parallax 
between the lens and source.  The event time scales were well measured for all events, 
and the Einstein radii were assessed for three of the four events.  In order to estimate the 
physical lens parameters with the constraints provided by the measured observables of the 
individual events, we conduct Bayesian analyses using a Galactic model.

\begin{figure}[t]
\includegraphics[width=\columnwidth]{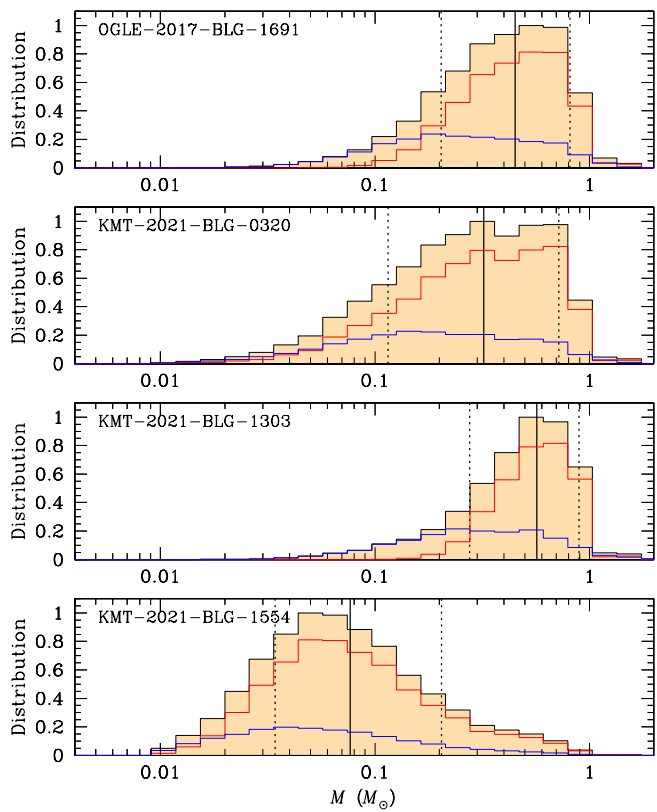}
\caption{
Bayesian posteriors of the host masses of the planetary systems. For each distribution, the 
solid vertical line indicates the median value, and the dotted lines represent 1$\sigma$ range 
of the distribution.  The blue and red curves represent the distributions contributed by disk 
and bulge populations of lenses, respectively, and the black curve is the sum from the two lens 
populations.
}
\label{fig:eleven}
\end{figure}

In the Bayesian analysis, we started by producing many lensing events from a Monte Carlo 
simulation conducted with the use of a Galactic model, which defines the matter-density and 
kinetic distributions, and mass function of Galactic objects.  We adopted the Galactic model 
constructed by \citet{Jung2021}.  In this model, the density distributions of disk and bulge 
objects are described by the \citet{Robin2003} and \citet{Han2003} models, respectively.  The 
kinematic distribution of disk objects is based on the modified version of the \citet{Han1995} 
model, in which the original version is modified to reconcile the changed density model, that 
is, the \citet{Robin2003} model.  The bulge kinematic distribution is modeled based on proper 
motions of stars in the Gaia catalog \citep{Gaia2016, Gaia2018}.  For the details of the density 
and kinematic distributions, see \citet{Jung2021}.  In the Galactic model, the mass function is 
constructed with the adoption of the initial mass function and the present-day mass function of 
\citet{Chabrier2003} for the bulge and disk lens populations, respectively.

\begin{figure}[t]
\includegraphics[width=\columnwidth]{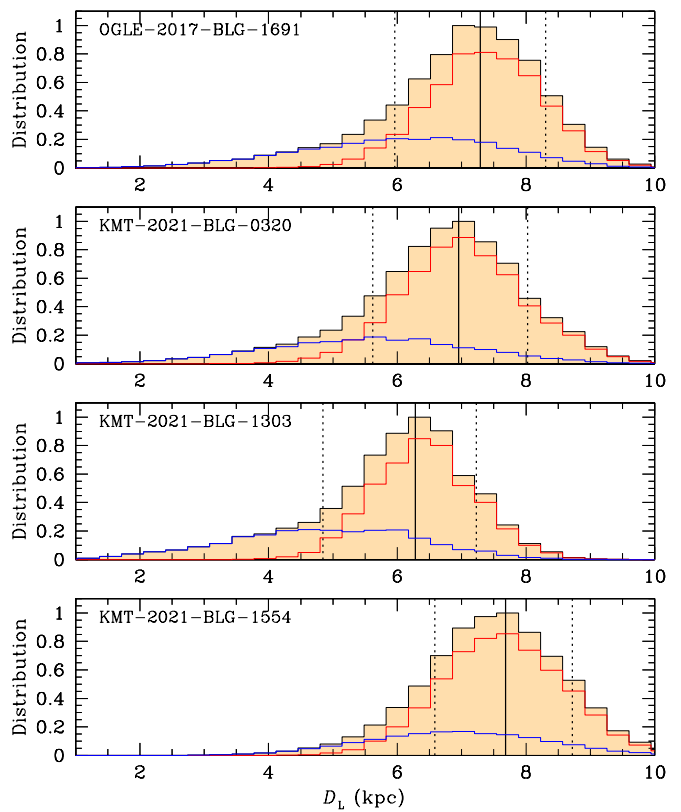}
\caption{
Bayesian posteriors of the distances to the
planetary systems. Notations are same as those in
Fig.~\ref{fig:eleven}.
}
\label{fig:twelve}
\end{figure}

In the second step, we computed the time scales and Einstein radii of the 
simulated events using the relations in Equation~(\ref{eq6}), and then constructed posterior 
distributions of $M$ and $\dl$ for the simulated events with the $\te$ and $\thetae$ values that 
are consistent with the observables of the individual lensing events. For the three events 
OGLE-2017-BLG-1691, KMT-2021-BLG-1303, and KMT-2021-BLG-1554, we use the constraints of both 
$\te$ and $\thetae$, and for KMT-2021-BLG-0320, we apply the constraint of only $\te$ because 
$\thetae$ is not measured for the event.

Figures~\ref{fig:eleven} and \ref{fig:twelve} show the Bayesian posteriors of $M$ and $\dl$ 
of the individual events, respectively.  In Table~\ref{table:nine}, we summarize the estimated 
values of the host and planet masses, $M_{\rm host}$ and $M_{\rm planet}$, distance, and projected 
host-planet separation, $a_\perp=s\dl \thetae$, for all of which medians are presented as 
representative values and the uncertainties are estimated as 16\% and 84\% of the Bayesian 
posterior distributions.  The median values and uncertainty ranges of the individual events 
are marked by solid and dotted vertical lines in the Bayesian posteriors presented in 
Figures~\ref{fig:eleven} and \ref{fig:twelve}, respectively.  For the events with degenerate 
solutions, OGLE-2017-BLG-1691, KMT-2021-BLG-0320 and KMT-2021-BLG-1554, we present the planetary 
separations $a_\perp$ corresponding to both the close and wide (or inner and outer) solutions.
We note that the uncertainties of the estimated physical parameters are big because of the 
intrinsically weak Bayesian constraint.  For example, from the comparison of the posterior 
distributions of OGLE-2017-BLG-1691 and KMT-2021-BLG-0320, one finds that the difference 
between the two posterior distributions is not significant, although the mean lens mass 
for the former event is slightly bigger than that of the latter event due to the longer 
time scale and the uncertainty is smaller due to the additional constrain of $\thetae$.

The estimated masses indicate that all of the discovered planets have sub-Jovian masses.  The 
planet masses KMT-2021-BLG-0320L, KMT-2021-BLG-1303L, and KMT-2021-BLG-1554L correspond to 
$\sim 0.10$, $\sim 0.38$, and $\sim 0.12$ times of the mass of the Jupiter, and the planet mass 
of OGLE-2017-BLG-1691L corresponds to that of Uranus.  To be noted among the planet hosts is that 
the estimated mass of the planetary system KMT-2021-BLG-1554L,  $M_{\rm host}\sim 0.08~M_\odot$, 
corresponds to the boundary between a star and a brown dwarf (BD).  Together with the previously 
detected planetary systems with very low-mass hosts\footnote{MOA-2011-BLG-262 \citep{Bennett2014},
OGLE-2012-BLG-0358 \citep{Han2013},
MOA-2015-BLG-337 \citep{Miyazaki2018},
OGLE-2015-BLG-1771 \citep{Zhang2020},
KMT-2016-BLG-1820 \citep{Jung2018a},
KMT-2016-BLG-2605 \citep{Ryu2021},  
OGLE-2017-BLG-1522 \citep{Jung2018b}, 
OGLE-2018-BLG-0677 \citep{Herrera2020},
KMT-2018-BLG-0748 \citep{Han2020b}, and
KMT-2019-BLG-1339L \citep{Han2020a}},
the discovery of the system demonstrates the microlensing capability of detecting planetary systems 
with very faint or substellar hosts.  Besides KMT-2021-BLG-1554L, the host stars of the other planetary 
systems are low-mass stars with masses lying in the range of $\sim [0.3$--$0.6]~M_\odot$.  Under the 
approximation that the snow line distance scales with the host mass as $a_{\rm sl}\sim 2.7(M/M_\odot)$~AU, 
all discovered planets lie beyond the snow lines of the hosts regardless of the solutions, demonstrating 
the high microlensing sensitivity to cold planets.  The planetary systems lie in the distance range of 
$\sim [6.3$--$7.6]$~kpc, demonstrating the usefulness of the microlensing method in detecting remote 
planets.

\section{Summary and conclusion}\label{sec:six}

We presented the analyses of four planetary microlensing events OGLE-2017-BLG-1691, 
KMT-2021-BLG-0320, KMT-2021-BLG-1303, and KMT-2021-BLG-1554.  The events share a common 
characteristic that the planetary signals appeared as anomalies with very short durations, 
ranging from a few hours to a couple of days, and they were clearly detected solely by the 
combined data of the high-cadence lensing surveys without additional data from followup 
observations.

From the detailed analyses of the events, it was found that the signals were generated by 
planets with low planet-to-host mass ratios: three of the planetary systems with mass ratios 
of the order of $10^{-4}$ and the other with a mass ratio slightly greater than $10^{-3}$.  
In the histogram of microlensing planets presented in Figure~\ref{fig:one}, we mark the 
positions of the four planets discovered in this work.  The estimated masses indicated that 
all discovered planets have sub-Jovian masses, in which the planet masses of KMT-2021-BLG-0320Lb, 
KMT-2021-BLG-1303Lb, and KMT-2021-BLG-1554Lb correspond to $\sim 0.10$, $\sim 0.38$, and $\sim 0.12$ 
times of the mass of the Jupiter, and the mass of OGLE-2017-BLG-1691Lb corresponds to that of the 
Uranus.  It was found that the host of the planetary system KMT-2021-BLG-1554L has a mass at around 
the boundary between a star and a brown dwarf.  Besides this system, it was found that the host 
stars of the other planetary systems are low-mass stars with masses in the range of 
$\sim [0.3$--$0.6]~M_\odot$.  The discoveries of the planets well demonstrate the capability of 
the current high-cadence microlensing surveys in detecting low-mass planets.

\begin{acknowledgements}
Work by C.H. was supported by the grants  of National Research Foundation of Korea 
(2020R1A4A2002885 and 2019R1A2C2085965).
This research has made use of the KMTNet system operated by the Korea Astronomy and Space 
Science Institute (KASI) and the data were obtained at three host sites of CTIO in Chile, 
SAAO in South Africa, and SSO in Australia.
The OGLE project has received funding from the National Science Centre, Poland, grant
MAESTRO 2014/14/A/ST9/00121 to AU.
he MOA project is supported by JSPS KAKENHI grant Nos. JSPS24253004,
JSPS26247023, JSPS23340064, JSPS15H00781, JP16H06287, JP17H02871, and JP19KK0082. 
\end{acknowledgements}

\end{document}